\documentclass[12pt]{article}
\usepackage[dvips]{graphicx}
\usepackage{color}
\usepackage{amssymb}
\usepackage{amsmath}
\usepackage{epsfig}

\usepackage{epsf}
\usepackage{graphicx,epsfig}
\usepackage{amsfonts}
\usepackage{amssymb}
\usepackage{cite}

\usepackage{euscript}
\usepackage{amssymb}
\usepackage{amsfonts}
\usepackage{amsbsy}
\usepackage{amsmath}
\usepackage{epsfig}
\usepackage{amsthm}
\usepackage{amscd}
\usepackage{amstext}
\usepackage{hyperref}

\def\gsim{\, \rlap{$>$}{\lower 1.1ex\hbox{$\sim$}}\,}
\def\lsim{\, \rlap{$<$}{\lower 1.1ex\hbox{$\sim$}}\,}

\makeatletter
\renewcommand\section{\@startsection {section}{1}{\z@}%
                                 {-3.5ex \@plus -1ex \@minus -.2ex}
                                   {2.3ex \@plus.2ex}%
                                   {\normalfont\large\bfseries}}
\renewcommand\subsection{\@startsection{subsection}{2}{\z@}%
                                   {-3.25ex\@plus -1ex \@minus -.2ex}%
                                     {1.5ex \@plus .2ex}%
                                     {\normalfont\bfseries}}
\renewcommand\subsubsection{\@startsection{subsubsection}{3}{\z@}%
                                   {-3.25ex\@plus -1ex \@minus -.2ex}%
                                     {1.5ex \@plus .2ex}%
                                     {\normalfont\itshape}}
\makeatother

\newcommand{\be}{\begin{equation}}
\newcommand{\ee}{\end{equation}}
\newcommand{\bea}{\begin{eqnarray}}
\newcommand{\eea}{\end{eqnarray}}
\newcommand{\barr}{\begin{array}}
\newcommand{\earr}{\end{array}}

\def\beq{\begin{equation}}
\def\eeq{\end{equation}}
\def\be{\begin{equation}}
\def\ee{\end{equation}}
\def\bea{\begin{eqnarray}}
\def\eea{\end{eqnarray}}

\DeclareRobustCommand{\SkipTocEntry}[4]{}

\begin{document}

\begin{titlepage}

\setcounter{page}{1} \baselineskip=15.5pt \thispagestyle{empty}

\begin{flushright}
SU-ITP-10/33 \\
SLAC-PUB-14305\\
NSF-KITP-10-141\\
DESY 10-201
\end{flushright}
\vfil

\begin{center}

{\Large \bf Simple exercises to flatten your potential}
\\[0.7cm]
{\large Xi Dong,$^{1,2}$ Bart Horn,$^{1,2}$  Eva Silverstein,$^{1,2}$ and Alexander Westphal$^{3,1}$}
\\[0.7cm]

{\normalsize {\sl $^{1}$ SLAC and Department of Physics, Stanford University, Stanford, CA 94305, USA}}\\

\vspace{.3cm}
{\normalsize { \sl $^{2}$ Kavli Institute for Theoretical Physics, University of California, Santa Barbara, CA 93106, USA}}\\

\vspace{.3cm}
{\normalsize { \sl $^{3}$ Deutsches Elektronen-Synchrotron DESY, Theory Group, D-22603 Hamburg, Germany}}\\

\end{center}

\vspace{.8cm}

\hrule \vspace{0.3cm}
{\small  \noindent \textbf{Abstract} \\[0.3cm]
\noindent
We show how backreaction of the inflaton potential energy on heavy scalar fields can flatten the inflationary potential, as the heavy fields adjust to their most energetically favorable configuration.  This mechanism operates in previous UV-complete examples of axion monodromy inflation -- flattening a would-be quadratic potential to one linear in the inflaton field -- but occurs more generally, and we illustrate the effect with several examples. Special choices of compactification minimizing backreaction may realize chaotic inflation with a quadratic potential, but we argue that a flatter potential such as power-law inflation $V(\phi) \propto \phi^p$ with $p<2$ is a more generic option at sufficiently large values of $\phi$.
}\vspace{0.5cm}  \hrule

\vfil
\begin{flushleft}
\today

\end{flushleft}

\end{titlepage}

\newpage
\tableofcontents
\newpage

\section{Motivation:  realizing your potential}

Inflation \cite{Inflation, Linde:1983gd} is a powerful framework for addressing the cosmological flatness and horizon puzzles, and for generating the primordial seeds of structure.
One recent advance is the development of a model-independent ``bottom-up" effective field theory framework \cite{Cheung:2007st}, which organizes CMB observables in terms of the lowest dimension operators participating in the effective theory. Still, model building plays an important role, both in field theory\footnote{See \cite{Baumann:2010nu}\ for a recent example.} and from the ``top down" in string theory.  In particular,
inflation is sensitive to Planck-suppressed higher dimension operators in the low energy Lagrangian (an infinite sequence of them in the case of large-field inflation with detectable tensor modes).
It is therefore of interest to model inflation within a UV-complete candidate for quantum gravity, of which string theory is our best-studied example (see \cite{Reviews}\ for a recent review).

The extra degrees of freedom of string theory -- arising at various mass scales up to the four-dimensional Planck scale -- affect the effective action along candidate inflaton directions in field space.
This has led to important constraints and complications, such as order one corrections to slow roll parameters from compactification effects \cite{Kachru:2003sx}\ and bounds on the inflationary energy relative to the scale of moduli stabilizing potential energy barriers \cite{KL}\cite{Silverstein:2008sg}.

Additional fields can play other roles, sometimes in fact contributing useful effects to model building.  The string theory motivated
possibility of many additional light fields assisting inflation has been addressed in works such as \cite{Nflation}, and the tendency of particle production to slow down the inflaton was analyzed in \cite{Green:2009ds}.\footnote{For similar approaches using a gas of particles to slow the inflaton field on a steep potential see e.g. \cite{Berera:2008ar,DK}.}
In some circumstances, integrating out heavy fields changes the character of the inflationary mechanism, producing higher dimension operators suppressed by the inflaton.  An early example of this is \cite{DBI}\ where off-diagonal Yang-Mills matrix fields renormalize the effective action for the diagonal fields.  In \cite{Tolley:2009fg}\ similar effects were constructed via integration out of heavy fields coupled through the kinetic term.  Integrating out heavy fields can also introduce a field-dependent enhancement of the kinetic term in the inflaton equation of motion \cite{Rubin:2001in}\ or produce features in the power spectrum for small enough radius of curvature in field space (see e.g. \cite{Patil}\ for a recent discussion).  Effects of heavy fields on precision observables such as the spectral tilt and the tensor to scalar ratio were considered in \cite{Bartolo:2007hx}.

In this note, we show how interactions with heavy scalar fields -- such as moduli and KK modes -- can help flatten the inflaton potential.  This mechanism was used in the small-field models of \cite{smoothhybrid} but can occur very generally.
The reason is very simple:  the heavy fields coupled to the inflaton relax to their most energetically favorable configuration.  Consider, as motivation, a simple field theoretic toy model with two fields $\phi_L$, $\phi_H$ with the following potential
\beq
V(\phi_L, \phi_H) = g^2 \phi^2_L \phi^2_H + m^2(\phi_H - \phi_0)^2~.
\eeq
The light field $\phi_L$ will play the role of the inflaton in this toy model.  Assuming its kinetic energy is a subdominant effect
(as we will shortly confirm),
the heavy field will track its instantaneous minimum, which is itself a function of $\phi_L$, and so the potential takes the form
\beq
V(\phi_L, \phi_{H,min}(\phi_L)) = \frac{g^2 \phi^2_L}{g^2 \phi^2_L + m^2} m^2 \phi^2_0~.
\eeq
For $\phi_L \gg m/g$, the inflationary potential is nearly flat.  The Friedmann equation becomes $3H^2M^2_P \approx m^2 \phi^2_0$, and
\beq
H^2 \sim m^2 \frac{\phi^2_0}{M_P^2}~.
\eeq
We take $\phi_0$ to satisfy $0 < \phi_0 \ll M_P$ so that $m\gg H$,
enforcing that $\phi_H$ be heavy enough not to produce scalar perturbations during inflation.\footnote{Such fluctuations from additional light fields are constrained by existing limits on isocurvature fluctuations and non-Gaussianities in the CMB.  \cite{Observations}}
As mentioned above, here we ignored the time derivative terms in the $\phi_H$ equation of motion.  The ratio between $3H\dot\phi_H$ and a typical term $\sim g^2\phi_H\phi_L^2 $ in $\partial_{\phi_H}V$ is tiny in our solution, of order $(m/g\phi_L)^4(\phi_0/\phi_L)^2$.

This mechanism can operate purely within field theory.  However string theory naturally provides a wealth of heavy scalar fields coming from moduli stabilization and from Kaluza-Klein modes which may play the role of $\phi_H$, as well as potentially lighter fields such as axions and certain brane positions that may play the role of the light inflaton $\phi_L$.
In a general compactification we expect couplings between axions, fluxes and geometry.  As long as the moduli are not destabilized in the process\footnote{Although this is a more energetically favorable outcome, it requires the fields to go over moduli-stabilizing barriers.} the adjustments of the heavy fields will generically go in the direction of flattening the potential.  (For restricted couplings, this can fail; for example if we shifted $\phi_0$ by a term proportional to $\phi_L$ in the above example, it becomes quadratic at large field values, and can even steepen to quartic for a finite range of $\phi_L$ depending on parameter choices.)

One interesting consequence of this concerns $m^2\phi^2$ chaotic inflation, a classic model \cite{Linde:1983gd}.  The couplings in the effective action including the light and heavy fields are analytic, and the scalar potential is generically quadratic around an extremum of the potential.  In string theory, a key example of such a quadratic term descends from couplings of the form $|B\wedge F|^2$ in the low energy effective action, where $B$ is a two-form potential field which produces an axion upon integration over a two-cycle in the compactification.  However, although the potential is quadratic near the origin, the response of the heavy fields generically flattens the potential further out.  The models of \cite{McAllister:2008hb}\ in which the potential ends up linear in $\phi_L$ for $\phi_L>M_P$ is a particular example of this.  The present work aims to provide a more systematic understanding of this theoretical trend. (See \cite{Kaloper:2008qs,Kaloper:2008fb}\ for an interesting discussion of $m^2\phi^2$ inflation from flux monodromy developed within an effective field theory framework.)

\begin{figure}[t!]
\begin{center}
\includegraphics[width=10cm]{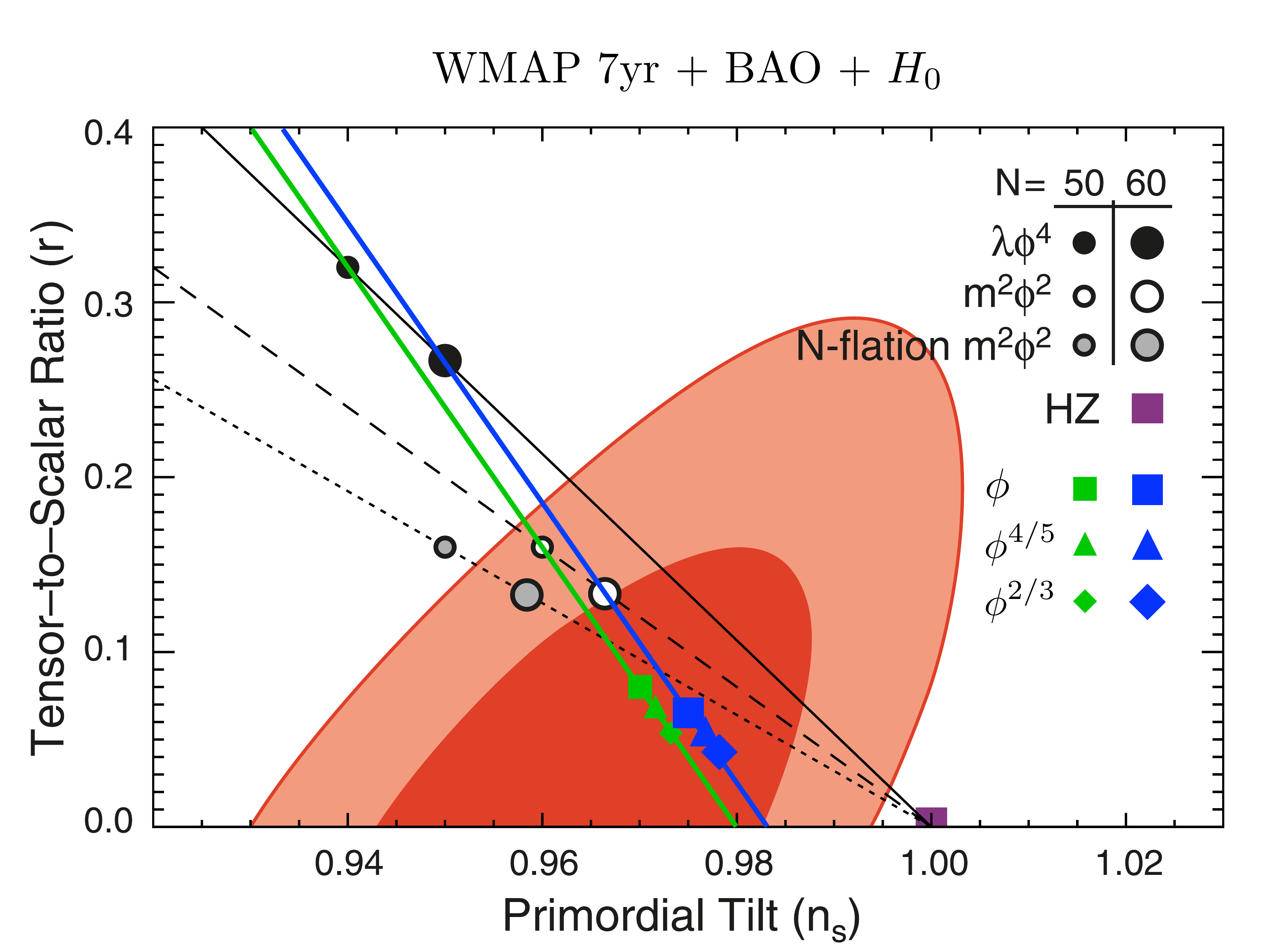}
\end{center}
\caption{Combined data constraints on the tensor to scalar ratio $r$ and the tilt $n_s$ \cite{Observations}\ together with the predictions for power-law potentials $\propto \phi^p\;,\;p>0$ for 50 e-foldings (green line) and 60 e-foldings (blue line) of inflation. Flattening the potential corresponds to moving down and to the right along these lines.  The colored points denote powers that have arisen in various large-field monodromy inflation models in string theory: IIB linear axion monodromy from 5-branes (squares; $\phi$), IIA moving 4-brane monodromy (diamonds; $\phi^{2/3}$), and a candidate example of IIB flux axion monodromy (this work; triangles; $\phi^{4/5}$).}
\label{fig:WMAP7}
\end{figure}

Observationally, a quadratic potential is still viable, currently sitting at the edge of the $1 \sigma$ exclusion contours, with smaller powers (corresponding to flatter potentials) lying further inside the allowed region \cite{Observations}.  Upcoming measurements \cite{Planck} are expected to significantly improve the constraints on the tensor to scalar ratio and the tilt of the power spectrum.
Because of the effects of heavy fields, including the flattening effect we consider here,
it would not be surprising if the $m^2\phi^2$ model gets excluded.
Special choices of compactification minimizing backreaction may realize chaotic inflation with a quadratic potential, but flatter potentials such as power-law inflation $V(\phi) \propto \phi^p$ with $p<2$ appear to arise more generically at sufficiently large values of $\phi$. We illustrate the predictions of a flattening monomial power-law potential against the present status of the WMAP 7-year results for the CMB in Fig.~\ref{fig:WMAP7}.

This paper is organized as follows.  In the remainder of this section and
the next, we introduce the general setup, further specify conditions under which
the energetic argument leading to flattening of the potential applies, and
describe important situations where it fails.  In section 3 we give several
distinct realizations of the effect in the context of axion inflation
in string theory, with different fields playing the role of $\phi_H$.    
In section 4 we make some concluding remarks.

\subsection{Additional kinetic effects}
\label{subsec:kineticconstraints}

In the toy model presented above, we solved for $\phi_H$ in terms of $\phi_L$ to good approximation by solving $\partial_{\phi_H}V(\phi_L,\phi_H)\equiv 0$; the kinetic term $\dot\phi_H^2$ was subdominant.  In more general examples we will need to establish whether the same approximation holds.  Consider an action (for homogeneous fields) of the form
\beq\label{moregen}
\int d^4 x\sqrt{-g}\left\{\dot\phi_H^2+G_{LL}(\phi_H)\dot\phi_L^2- V(\phi_L,\phi_H) \right\}~.
\eeq
In integrating out $\phi_H$, there are two effects that may arise from the kinetic terms.  The first, discussed in \cite{Rubin:2001in}, is that the $\dot\phi_H^2$ kinetic term affects the solution for $\phi_H$ as $\phi_L$ rolls.  This is significant if $|d\phi_H/d\phi_L|$ is large compared to $\sqrt{G_{LL}}$.  In our examples, as in the above toy model, we will check that this quantity is small.

The second, discussed in \cite{Tolley:2009fg}, arises from the coupling of $\phi_H$ in the light field's kinetic term $G_{LL}(\phi_H)\dot\phi_L^2$.  If $\dot\phi_L^2$ is large enough during inflation, this term can significantly affect the solution for $\phi_H$, leading to a nontrivial k-inflationary \cite{ArmendarizPicon:1999rj}\ effective Lagrangian ${\cal L}[(\partial\phi_L)^2,\phi_L]$.  In this class of models, inflation may occur on a steep potential, with self-interactions of the field $\phi_L$ slowing it down (resulting in a large non-Gaussian signature in the power spectrum).  The energetics of the backreaction for these more general solutions is not as simple as it is in the limit of slow roll inflation, where the heavy fields adjust in such a way as to flatten the potential when possible.
Within slow roll inflation we have $\dot\phi_L^2\ll V$, and this will allow us to self-consistently bound the effect in our examples below. It would be interesting to find UV-complete examples of the effects in \cite{Tolley:2009fg,Rubin:2001in} in future work.

\subsubsection{Steepening from kinetic curvature}

We should emphasize that flattening of the potential is not an automatic consequence of couplings to massive fields.  For example, even
when the kinetic effects of the previous subsection are small it can fail, as can be seen from the following variant of our previous toy model:
\beq
\mathcal{L} = \frac{1}{2}\frac{\phi_H}{M_P}\dot\phi_L^2 + \frac{1}{2}\dot\phi_H^2 - g^2 \phi^2_L \phi^2_H - m^2(\phi_H - \phi_0)^2 - \mu^2 \phi_L^2~.
\eeq
\noindent As before, for large $\phi_L,$ $|d\phi_H/d\phi_L|$ is small compared to $\sqrt{G_{LL}}$ and $\dot\phi^2_H$ can safely be neglected, and the kinetic term $\frac{\phi_H}{M_P} \dot\phi^2_L$ is subdominant to the potential, so that the effects of \cite{Tolley:2009fg} are suppressed.  The canonical field $\tilde{\phi}$ at large $\phi_L$ is now $\approx \frac{m}{g}\sqrt{\frac{\phi_0}{M_P}}$ log $(\phi_L/M_P),$ and the potential has the form
\beq
V_{eff}(\tilde{\phi}) \approx m^2 \phi^2_0 + \mu^2 M^2_P e^{\frac{2g\tilde{\phi}}{m}\sqrt{\frac{M_P}{\phi_0}}}~.
\eeq
\noindent Thus, if $G_{LL}(\phi_{H,min}(\phi_L))$ scales like a negative power of $\phi_L,$ then the dressed kinetic term is responsible for \textit{steepening} the potential. This inverse power can arise for example in the case where $\phi_H$ descends from the overall (inverse) volume of a string compactification, leading to an increased volume at large inflaton field values (fattening the manifold, and steepening the potential).   This can be neglected in examples with sufficiently strong volume-stabilizing potential barriers. In a complete example, $\mu$ would likely not be a fixed parameter, and all backreaction effects would need to be incorporated consistently.

\section{Warmup:  review of axion monodromy inflation}

Our string-theoretic examples grew out of a project aimed at developing the flux version of axion monodromy inflation.
Let us begin by briefly reviewing the general discussion of this mechanism in \cite{McAllister:2008hb}.
A flux version of monodromy inflation has been obtained at the level of effective field theory also in \cite{Kaloper:2008fb,Kaloper:2008qs}, and the phenomenology of monodromy inflation was further developed in \cite{Flauger:2009ab,BP}.

String theory naturally provides axions
\beq\label{axion}
b=\int_{\Sigma^p}B_p~~,~~ c=\int_{\Sigma^p}C_p
\eeq
coming from p-form Neveu-Schwarz--Neveu-Schwarz and Ramond-Ramond fields $B_p$, $C_p$ wrapped on p-cycles in the compact directions.
Assuming a single scale $L\sqrt{\alpha'}$ for the compactification geometry, the canonically normalized field is related to the angular scalar field (\ref{axion}) by
\beq
\frac{\phi_b}{M_p} \sim \frac{b}{L^p}~~,~~ \frac{\phi_c}{M_p} \sim \frac{g_s c}{L^p}~.
\eeq
The theory contains couplings between the axions and various fluxes and spacefilling branes that are generically present in compactifications.  These couplings introduce {\it monodromy} in the axion direction:  the system builds up potential energy as $b$ or $c$ traverses its basic period.

In the specific, UV-complete examples discussed in \cite{McAllister:2008hb} the axion potential is lifted by the DBI action
\beq\label{DBI}
{\cal S}_{DBI}=-\frac{1}{g_s \alpha'^{3}}\int \sqrt{det(G_{MN}+B_{MN})\partial_\alpha X^M\partial_\beta X^N}~~\Rightarrow~~
V(\phi_b)\propto \sqrt{1+\left(\frac{\phi_b}{M_P}\right)^2}
\eeq
for a spacefilling D5-brane wrapped on the 2-cycle (or its S-dual in the case of RR axions).
Using the AdS/CFT correspondence, this result can be described equivalently in terms of a dual geometry plus fluxes. In that description, the monodromy arises from flux couplings of the form
\beq\label{flux}
{\cal L}\sim |B_2\wedge F_3|^2 + \dots
\eeq
or its S-dual $|C_2\wedge H_3|^2$.  Although the coupling (\ref{flux}) is quadratic, backreaction of the axion and fluxes on the geometry leads to a linear potential, as we will discuss in more detail below.
This provides an explicit example of the general trend discussed in the introduction: that back reaction of the potential energy descending from (\ref{flux}) should flatten the potential, since this is energetically favorable.

Globally, however, the most energetically favorable configuration in metastable string compactifications is the runaway to large radius and/or weak coupling, or decays to negative cosmological constant.  Therefore,
before discussing examples of potential-flattening effects, let us first briefly review the combined conditions for maintaining moduli stabilization and the COBE normalization of the power spectrum.

As is emphasized in \cite{McAllister:2008hb}, the canonically normalized axion potential is, in the absence of strong warping (supposing for illustration that $B_2$ is the inflationary axion)
\beq\label{fluxenI}
\frac{1}{{\alpha'}^4}\int d^6 x\sqrt{-g}|B_2 \wedge F_q|^2 \sim \frac{1}{{\alpha'}^4}\frac{\phi^2_b}{M^2_P}\int d^6 x\sqrt{-g} |F_q|^2~.
\eeq
If the $q$-form flux lifting the axion potential makes a sufficiently subleading contribution to the moduli stabilization, one can obtain a super-Planckian field range without destabilizing the moduli.

In order to provide a successful phenomenological model of chaotic inflation, we must have a sufficient range to give $N_e = 60$ e-foldings of inflation, and the power spectrum of scalar perturbations must match the COBE normalization,
\beq
\Delta^2_{scalar} = \frac{H^4}{(2\pi)^2\dot{\phi}^2} \cong 10^{-9}~.
\eeq
For a power-law potential $V(\phi) \propto \mu^{4-n}\phi^n,$ the required field range is $\Delta \phi / M_p \sim \sqrt{n N_e},$ which is $O(15)$ for the quadratic case.  The COBE normalization becomes
\beq
\left(\frac{\mu}{M_P}\right)^{2-\frac{n}{2}}\left(\frac{\Delta \phi}{M_P}\right)^{\frac{n}{2}+1} \sim 10^{-5}
\eeq
\noindent which becomes $\mu / M_P \sim 10^{-6}$ for a quadratic potential and $O(10^{-3})$ for a linear potential.

Let us first review the basic scales in the problem which show that it is possible for axion monodromy inflation to self-consistently satisfy the required number of e-foldings and COBE normalization. Here is an estimate of the effects of these observational constraints
in the extreme case of $m^2\phi^2$ inflation, in the absence of warping (flatter potentials and warped models being easier to embed below the moduli stabilizing barrier, this is the most conservative estimate we can make).  Supposing that the inflaton comes from a $C_p$ axion lifted by the term $|C_p \wedge H_3|^2,$ the flux potential is
\beq\label{fluxenII}
\mathcal{U} = \frac{1}{{\alpha'}^4}\int d^{6}x \sqrt{-g}|C_p \wedge H_3|^2 \sim \frac{M^2_P}{\alpha'} \left(\frac{g_s c}{L^p}\right)^2\left(\frac{K}{L^3}\right)^2 \sim M^4_P \left(\frac{g^2_s K^2}{L^{12}}\right)\frac{\phi^2_c}{M^2_P}
\eeq
\noindent where we have labeled the number of $H_3$ flux quanta by $K$. The condition for realizing 60 e-foldings of inflation without destabilizing the moduli and for matching the power spectrum to the COBE normalization then becomes (returning to the general case of $q$-form flux lifting the inflaton $\sim |C_{p}\wedge F_{q}|^{2}$)
\beq
15K_{inf}\ll K_{moduli}~~;~~\frac{\mu}{M_p} \cong \frac{g_s K_{inf}}{L^{q+3}}(2\pi)^{7/2} \cong 10^{-6}~.
\eeq
\noindent
These conditions can be satisfied for reasonable parameter values, e.g. $g_s \sim 0.02,$ $K_{inf} \sim 1,$ $q=3,$ $L \sim 10.$  Moreover, as already mentioned,
warping can naturally suppress the potential energy if the inflationary sector is localized in a region of large gravitational redshift, as in the specific examples in \cite{McAllister:2008hb}.
Therefore there is no immediate obstruction to fitting the flux-based version of axion monodromy inflation into stabilized string compactifications, avoiding catastrophic decay of the vacuum.

More generally, there may be single-sector models where the inflaton potential itself helps stabilize the moduli during inflation, competing with or even dominating over some of the terms in the moduli potential.  The gravity dual of the models \cite{McAllister:2008hb}\ is a familiar local example of this, where down the brane throat the axion $c=\int_{\Sigma_2} C_2$ helps stabilize the cycle $\Sigma_2$ it threads.  Below we will explore potential generalizations of this which are further from a simple brane construction.

\subsection{Flattening vs. moduli potential barriers}

Before proceeding to our main flattening exercises, it is worth describing a simple example which illustrates both the flattening effect and how the requirement of moduli stabilization can cut it off.
De Sitter vacua can plausibly be achieved in string theory via perturbative techniques, where localized sources of energy such as curvature, D-branes and NS5-branes, fluxes, orientifolds and others contribute to an effective potential for the four dimensional scalar fields, which is minimized to solve the equations of motion.  Such constructions were introduced in \cite{Silverstein:2004id}\ and discussed in \cite{classicaldS}; worked examples include \cite{Maloney:2002rr,Saltman:2004jh, Silverstein:2007ac, Dong:2010pm}.  It is useful to organize these mechanisms in terms of an `abc' structure for the potential,
\beq
V(g)=ag^2-bg^3+cg^4
\eeq
where $g$ is a representative modulus such as the the string coupling (with the coefficients $a$, $b$, and $c$ depending on the other moduli).
Such a potential generally arises with curvature, Neveu-Schwarz--Neveu-Schwarz fluxes, and/or supercritical dimensionality in the $a$ term, orientifold planes in the $b$ term, and Ramond-Ramond fluxes in the $c$ term.
This potential has a positive metastable minimum when the quantity $4ac/b^2$ is minimized as a function of the other moduli, within the window
\beq
1 < \frac{4ac}{b^2} < \frac{9}{8}~.
\eeq
Adding flux energy from an axion term will produce an effective potential of the form
\beq
V(g,x)=ag^2-bg^3+(1+x^2)cg^4
\eeq
where $x$ is proportional to the axion field.  Explicit examples may be found among the axions in \cite{Saltman:2004jh, Silverstein:2007ac, Dong:2010pm}, though we have not developed complete models.

\begin{figure}[h]
\begin{center}
\includegraphics{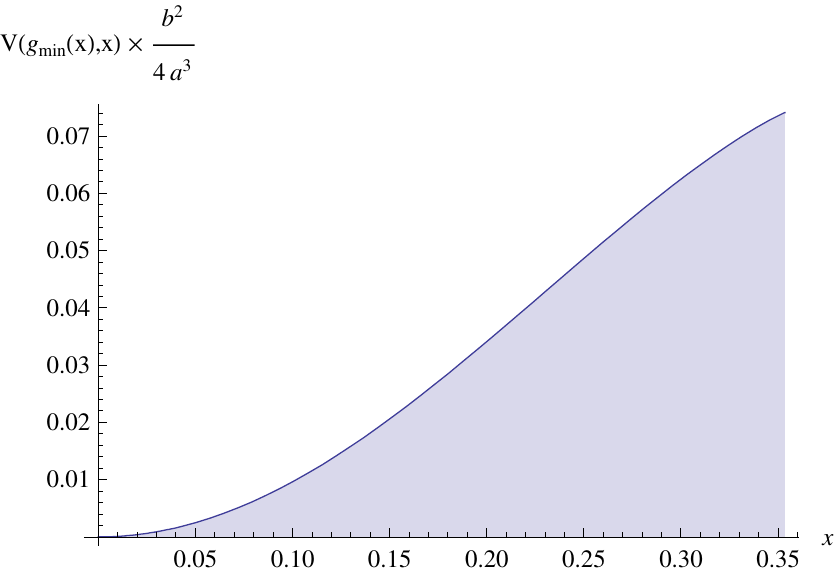}
\end{center}
\caption{Effects of an inflationary flux on the three-term structure stabilized in a Minkowski minimum for $x = 0$.}
\label{fig:Vmin}
\end{figure}

Setting $4ac/b^2 = 1$, the potential is stabilized at a Minkowski minimum for $x=0$, and as $x$ is turned on, the de Sitter minimum persists as long as
\beq\label{xbound}
x^2 < \frac{1}{8}~.
\eeq
Including backreaction, $V(g_{min}(x),x)$ is no longer quadratic, as plotted in Figure~\ref{fig:Vmin}.
As expected, the potential is quadratic for small values of $x$ where backreaction can be ignored, and then flattens as $x$ is increases.  However,
the flattening only starts to become significant when $x$ is of order one, but from (\ref{xbound}) it is clear that $x$  begins to destabilize the minimum at this point.

\section{Workout:  axions pushing on heavy fields}

Finally let us turn to the effects of interest in this paper, the backreaction of the energy (\ref{fluxenI})(\ref{fluxenII}) on heavy fields and its effect on the inflationary potential energy.

\subsection{Bowflux:  Sloshing of flux on fixed cycles}

The axion potential may be modified by rearrangement of fluxes on fixed cycles so as to minimize their energy.  To illustrate this effect, we consider a model of the kind discussed by \cite{Giddings:2001yu}\ stabilized by three-form fluxes $H_3$, $F_3$.  We add a small extra three-form flux $\Delta H_3$ to an unwrapped cycle, and turn on an axion $C_2$ threading a cycle $\Sigma_2$ as the inflaton.  Our candidate inflaton will be $c\sim \int_{\Sigma_2}C_2$.  We  minimize the other fields at a given value of $c$, given consistency with moduli stabilization which requires that the inflationary energy stay below the moduli-stabilizing barriers.
In general the potential will descend from terms in the 10D action of the form
\beq\label{sloshterms}
\frac1{g_s^2}|H_3+\Delta H_3|^2+|C_2\wedge (H_3+\Delta H_3)|^2+|F_3|^2~.
\eeq
The number of flux quanta threading a given cycle is topological and does not change, but the fluxes may slosh around on their cycles so as to minimize the total energy.    If the flux $\Delta H_3$ shifts so that its support is partially separated from that of $C_2$, for instance, the Chern-Simons term would be weakened, but the contribution to the potential from the $|\Delta H_3|^2$ term would increase.  The competition between them determines the optimal field configuration.
In general the geometry and the axion wavefunction can adjust as well. Before considering the potential energy, $C_2$ minimizes its energy by forming a flat connection $c\omega_2$ (where $\omega_2$ is a nontrivial closed form which integrates to one over $\Sigma_2$).  In the presence of potential energy, it might prove energetically favorable for Kaluza-Klein modes of $C_2$ to turn on to reduce the second term in (\ref{sloshterms}), at the cost of introducing a contribution to the $|F_3|^2$ term. However,
to illustrate our effect, let us focus on the sloshing of $\Delta H_3$ at fixed $C_2$, since the adjustment of any other modes (such as the geometry and $C_2$ itself) can only enhance the flattening effect.

Keeping fixed the integral of $\Delta H_3$ over the three-cycles it threads, $\Delta H_3$ can scrunch up in three directions $w$ along the three-cycle to reduce its overlap with $C_2$.  Let us denote by $\tilde{L}\sqrt{\alpha'}$ the size of the region over which the scrunched-up field $\Delta H_3$ has support, modeling its profile locally by
\beq\label{Hgauss}
\sqrt{\alpha'}\Delta H_3\sim \frac{\Delta N}{{\tilde{L}}^3}e^{-{w^2/\tilde{L}^2\alpha'}}
\eeq
where $\Delta N$ is the number of $\Delta H_3$ flux quanta. We would like to minimize the potential energy with respect to $\tilde{L}$ and determine the effect of this on the axion potential.
If the profile of $C_2$ were flat in the internal dimensions, shrinking $\tilde{L}$ would not be advantageous.  Of course harmonic forms in nontrivial compactification manifolds are not constant.  Taylor expanding (and assuming rough isotropy locally), let us model $C_2$ in the region of support of $\Delta H_3$ as
\beq\label{Ctwo}
C_2(w)\sim \frac{c}{L^2}\left(1+\gamma \frac{{w}^2}{L^2\alpha'}+\dots\right)
\eeq
where $\gamma$ is a constant derived from the Taylor expansion of $C_2$'s profile.\footnote{Note that $C_2 \wedge H_3$ will in general include angular factors depending on the geometry.  We will not write these factors here.}
Here we are assuming $\Delta H_3$ is centered on a local minimum of $C_2$, which is its preferred configuration if available (otherwise one would obtain a linear term in the expansion (\ref{Ctwo}), with similar results).

After integrating over the internal volume, the relevant terms in the potential are proportional to
\beq\label{Bowfluxpot}
\frac1{g_s^2}\left[\Delta N^2\left(\frac{L}{\tilde{L}}\right)^3+\gamma N\Delta N\left(\frac{\tilde{L}}{L}\right)^2\frac{\phi^2_c}{M^2_P}\right]
\eeq
where $N$ refers to the number of $H_3$ flux quanta and $L\sqrt{\alpha'}$ is a typical length scale in the compactification. Here the first term comes from $|\Delta H_3|^2$.  The second term comes from $(C_2\wedge H_3)\cdot(C_2\wedge\Delta H_3)$, and gets its leading contribution from the ${w}^2$ term in (\ref{Ctwo}) convolved with (\ref{Hgauss}).\footnote{The other terms are either subleading or do not depend on $\tilde{L}$.}  The potential will then minimize these two terms and be proportional to $\phi_c^{6/5}$, which is flatter than quadratic.

This illustrates the flattening mechanism, but only provides a lower bound on the effect.  Adjustments of other fields including $C_2$ and the compactification geometry would further flatten the potential.

\subsubsection{Bounding additional kinetic effects}

As discussed above in \S\ref{subsec:kineticconstraints}, we must check whether it is a good approximation to determine the heavy field $\phi_H$ (in this case corresponding to KK modes of $B_2$) in terms of $\phi_L$ by solving $\partial_{\phi_H}V\equiv 0$, neglecting the contributions from the kinetic terms.  The kinetic effects of \cite{Rubin:2001in}\ are small if $|\partial\phi_H/\partial\phi_L|\ll 1$, i.e. if the KK modes of $B_2$ that we consider make a negligible correction to $C_2$'s kinetic term. It is straightforward to see that this may be obtained in the present example, as follows.
The kinetic term for $\tilde{L}$ descends from the kinetic term for $\Delta H_3$ and is
\beq
\mathcal L_{kin} \sim \frac{\Delta N^2 M_P^2}{{\tilde{L}}^3L^3} (\partial \tilde{L})^2~,
\eeq
giving the canonically normalized field $\phi_H$ as
\beq
\phi_H \sim \frac{\Delta N M_P}{{\tilde{L}}^{1/2}L^{3/2}} + const~.
\eeq
Minimizing the potential (\ref{Bowfluxpot}) with respect to $L'$, we get
\beq
\frac{L}{\tilde{L}} \sim \left(\frac{\gamma N}{\Delta N}\right)^{1/5} \left(\frac{\phi_c}{M_P}\right)^{2/5}~.
\eeq
Combining the above equations and writing $\phi_H$ as a function of $\phi_c$, we get
\beq
\frac{d\phi_H}{d\phi_c} \sim \frac{\Delta N}{L^2} \left(\frac{\gamma N}{\Delta N}\right)^{1/10} \left(\frac{M_P}{\phi_c}\right)^{4/5}~,
\eeq
which can be much smaller than 1 for a reasonable range of parameters.

\subsection{Puffing on the kinetic term}
\label{subsec:kinetic}

In the previous subsection we have considered modification of the effective potential due to backreaction on the potential terms.  The backreaction of the inflationary potential on the geometry can also affect the kinetic term, realizing the ``running kinetic term" mechanism described in \cite{Nakayama:2010kt}.

In a simple situation where the NS-NS or R-R field threads a cycle of size $L\sqrt{\alpha'}$ that is the same as the typical length scale in the compactification, the canonical normalization of the inflaton field is given in terms of the number of axion windings by
\beq
\frac{\phi_b}{M_P} \sim \frac{b}{L^2}~~,~~\frac{\phi_c}{M_P} \sim \frac{g_s c}{L^p}
\eeq
\noindent respectively for a 2-form NS-NS and for a p-form R-R axion.  If instead we consider cases where the p-form field is localized (e.g.\ in a throat) and is therefore threading a much smaller cycle of size $L'\sqrt{\alpha'}$, the canonically normalized field becomes
\beq
\frac{\phi_b}{M_P} \sim \frac{b}{L'^2}\frac{L'^3}{L^3} \sim \frac{b L'}{L^3}~~,~~\frac{\phi_c}{M_P} \sim \frac{g_s c}{L'^p}\frac{L'^3}{L^3} \sim \frac{g_s c L'^{3-p}}{L^3}~.
\eeq
Here we are considering the case that the support of the axion is of order the size $L'\sqrt{\alpha'}$ in all directions in the compactification (as occurs for example in the case that $L'\sqrt{\alpha'}$ describes the size of an internal cycle localized within a Freund-Rubin throat).
Now if the inflationary flux backreacts on the size $L'\sqrt{\alpha'}$ of the wrapped cycle, $L'$ will become a function of the axion and this will alter the relation between $b$ or $c$ and the canonically normalized field.  The terms of the form $|\textnormal{axion}\wedge\textnormal{flux}|^2$ push the geometry to expand.  Given this, $L'(b)$ will vary as a positive power of $b$ and reduce the power of $\phi_b$ in the potential. For example, in the case where the size $L'\sqrt{\alpha'}$ is mostly supported by $|B_2\wedge F_3|^2$, we have $L'^4\propto b$ and therefore $\phi_b\propto b^{5/4}$.  In the case where the inflation arises from a Ramond-Ramond field, we will have $p \leq 3$ for magnetic fluxes in six compact dimensions, and so $L'(c)$ will either reduce the power of the potential or leave it unchanged.

\subsubsection{Bounding additional kinetic effects}
\label{subsubsec:kin}

In this example, we solved for the heavy field $L'$ in terms of the light field $\phi_b$ (or $\phi_c$) by minimizing the potential in the $L'$ direction.  Let us now address the question of additional kinetic effects described in \S\ref{subsec:kineticconstraints}\ in the context of the present model.  Before describing the kinetic interactions of $\phi_b$ and $L'$, let us note that the overall size $L\sqrt{\alpha'}$ of the compactification will not be pushed far in the process given a sufficient hierarchy between the inflationary energy and the moduli stabilizing barriers.

First, let us check whether $|d\phi_H/d\phi_L|$ is small.  This requires knowledge of the kinetic term for $\phi_H$, i.e. the relation between the canonically normalized field $\phi_H$ and the modulus $L'$.
The kinetic term for $L'$ descends from the ten-dimensional Einstein term, and in four-dimensional Einstein frame is given by
\beq\label{Lprimekin}
\int d^4 x\sqrt{-g}M_P^2\left(\frac{L'}{L}\right)^6 \left(\frac{\partial L'}{L'}\right)^2
\eeq
in the above example. From this, the canonically normalized field $\phi_H$ is
\beq\label{phiheavy}
\phi_H\sim M_P \left(\frac{L'}{L}\right)^3~.
\eeq
Now, from the above-mentioned scaling $L'^4\propto b,\phi_b\propto b^{5/4}$, we obtain $\phi_H\propto \phi_b^{3/5}$ and \beq\label{HL}
\left| \frac{d\phi_H}{d\phi_L} \right|\propto \phi_L^{-2/5}
\eeq
which is $\ll 1$ for sufficiently large $\phi_b=\phi_L$.

Next, let us check that the kinetic term for $\phi_L=\phi_b$ does not constitute a significant source for $\phi_H$ as considered in \cite{Tolley:2009fg}.  To do this, write $L'\equiv L'_0 e^{\sigma'(t)/M_P}$ (note here $\sigma'$ is not the canonically normalized field).  The relevant terms in the effective action have the form
\beq\label{terms}
\int d^4 x\sqrt{-g}\left(e^{2\sigma'/M_p}\dot\phi_b^2-V(\sigma',\phi_b)\right)~.
\eeq
Each term in the potential scales like a power of $L'\propto e^{\sigma'/M_P}$.  Varying this action with respect to $\sigma'$, the first term is of order $\dot\phi_b^2/M_P$, much smaller than the second term which is of order $V/M_p$ during inflation.  Thus we can self-consistently ignore the effect of \cite{Tolley:2009fg}\ here.

\subsection{Weight lifting:  pushing on moduli}
\label{subsec:modulipush}

The fact that the axion $\times$ flux energy pushes on the moduli can lead to a similar but distinct effect from the backreaction on the inflaton kinetic term just discussed.
One concrete example of this is simply the one developed in \cite{McAllister:2008hb}, described in terms of its gravity dual.  Again, the term $|C_2\wedge H_3|^2$ is quadratic in the axion $c=\int C_2$.  But the axion builds up effective D3-brane charge, and from that point of view the potential should be linear in $cg_s$, which is proportional to the effective number of D3-branes. This works out because the generalized 5-form RR flux $\tilde F_5=C_2\wedge H_3+\dots$ backreacts on the moduli, giving a near horizon internal geometry with size $R\sqrt{\alpha'}$ depending on $c$ as
\beq\label{RN}
R^4\sim g_s\tilde N \sim g_s c\int_{S^3}H_3
\eeq
as in standard Freund-Rubin solutions.  Folding this into the effective action, we see that it scales like
\beq\label{actionbr}
{\cal S}\sim \frac{1}{\alpha'^4}\int d^{10}x\sqrt{-G}|\tilde F_5|^2+\dots \sim Vol(4d)\frac{{\tilde N}^2}{R^{10}}\times R^6\sim \frac{\tilde N}{g_s}Vol(4d)
\eeq
as befits a set of D3-branes (here $Vol(4d)$ is the volume of the worldvolume swept out by the brane).  A straightforward calculation of the four dimensional effective potential, derived for general warping in  \cite{Giddings:2005ff}\cite{Douglas:2009zn}, allows one to reproduce from the gravity side the corresponding four-dimensional Einstein frame potential energy $V$ descending from the brane throat:
\beq\label{Effpot}
V(c)\sim M_P^4 \left(\frac{g_s^2}{Vol}\right)^2 \frac{\tilde N}{g_s}
\eeq
where $Vol$ is the compactification volume in string units.
In the specific construction \cite{McAllister:2008hb}, the kinetic term of the axion was dominated by the ultraviolet region of the compactification well outside the brane throat.  Therefore, in that example the kinetic backreaction of \S\ref{subsec:kinetic}\ does not apply, but backreaction on the geometry (specifically, on the internal size $R\sqrt{\alpha'}$) flattens the potential from quadratic to linear.
In this example, the kinetic effects are bounded much as in \S\ref{subsubsec:kin}.

\subsection{Circuit training:  toward more generic UV complete examples}

A general string compactification involves multiple backreaction effects that are simultaneously important. We have not fully controlled any such example in this paper, but will note here an interesting candidate.
Consider an $S^3$ localized down a warped throat.  Put $M$ units of RR $F_3$ flux on its dual cycle $\tilde S^3$.  On the $S^3$ itself, put zero total units of flux, but introduce a topologically trivial configuration of $h$ units of $H_3=dB_2$ on one hemisphere (north of the equator, say) and $-h$ units on the other (south of the equator).  This will dynamically relax back down to zero, and if the geometry were fixed the $|H_3|^2$ term would produce a quadratic potential for the integral $b\equiv\int_{equator} B_2=h$ of $B_2$ over the equator of the $S^3$.  Backreaction, however, will change this significantly.  Consider starting the system in a configuration in which each hemisphere times the $\tilde S^3$ with flux is approximately solving the equations of motion as in \cite{Klebanov:2000hb}\cite{Giddings:2001yu}.  This constitutes, in effect, a 3-brane throat and an anti-3-brane throat at the bottom of the original throat. One can set this up explicitly in terms of two close-by conifold singularities with flux.  A similar construction with metastable fluxes on a noncompact Calabi-Yau geometry is studied in \cite{Aganagic:2006ex}.

Each throat carries potential energy of order $\tilde N\sim M b(t)$ including the backreaction of \S\ref{subsec:modulipush}.  Moreover, the kinetic energy of $b$ is subject to backreaction as in \S\ref{subsec:kinetic}.
The four-dimensional canonically normalized field $\phi_b$ in four-dimensional Einstein frame is given by
\begin{align}
\int d^4x \sqrt{-g_E}(\partial \phi_b)^2 &\sim \frac{1}{g_s^2 {\alpha'}^4}\int d^{10}x\sqrt{-g_{st}}(\partial B_2)^2 \sim \frac{1}{\alpha'}\int d^{4}x \sqrt{-g_{st}}\left(\frac{R^6}{g_s^2}\right)\frac{(\partial b)^2}{R^4} \nonumber\\
&\sim \int d^4x \sqrt{-g_E} M^2_P\left(\frac{g_s^2}{Vol}\right)\left(\frac{R^2}{g_s^2}\right)(\partial b)^2 \sim \int d^4x \sqrt{-g_E} \left(\frac{M^2_P}{Vol}\right)R^2(\partial b)^2 \nonumber\\
&\sim \int d^4x \sqrt{-g_E}\frac{M^2_P(g_s M)^{1/2}}{Vol}b^{1/2}(\partial b)^2
\end{align}
and so $\frac{\phi_b}{M_P} \sim \frac{(g_s M)^{1/4}}{\sqrt{Vol}}b^{5/4}$.
These two effects, taken together, suggest a potential
\beq\label{fourfifths}
V(\phi_b)=\mu^{16/5}\phi_b^{4/5}~.
\eeq
However, in order to obtain a concrete prediction for the evolution of this system, we would
require a better understanding of the region between the brane and antibrane throats and full control over all sources of backreaction in all directions in field space.  This would be interesting to pursue further.

\section{Cooldown}

A quadratic inflaton potential may be the simplest possibility from a bottom-up approach, but interactions with heavier fields typically deform the effective action, flattening the potential in the cases discussed here for a simple energetic reason.  This is a basic aspect of the UV sensitivity of inflation, complementary to others much discussed in the recent literature.  If the upcoming round of CMB measurements become consistent with the predictions of $m^2\phi^2$ chaotic inflation, this would significantly constrain the inflaton's couplings to additional fields, including those much heavier than the inflationary Hubble scale.    Conversely, if the mild trend in the data toward flatter potentials sharpens, the considerations of this paper may help explain the results.

In the case of axion monodromy inflation, we have outlined two specific mechanisms for backreaction to flatten the axion potential; in general the fluxes and the geometry will seek out the state of lowest energy consistent with the higher dimensional equations of motion.  In general, determining the correct form of the potential seems a complicated task.  Complete catalogs of the modes found in compactification geometries, such as \cite{Baumann:2009qx,Baumann:2010sx}, may be of use in constructing more explicit examples.  It would also be interesting to see if these considerations apply to other mechanisms for inflation, including general small field models and models with more generic kinetic terms where the energetic analysis is somewhat more complicated.\footnote{It would also be interesting to study backreaction further in models where a cycle size modulus plays the role of the inflaton such as \cite{Bond:2006nc}\ or \cite{Cicoli:2008gp}.}

\section*{Acknowledgements}

We thank Gary Horowitz, Hans Jockers, Shamit Kachru, Nemanja Kaloper, Albion Lawrence, Louis LeBlond, Juan Maldacena, Liam McAllister, David Morrison, Matt Roberts, Leonardo Senatore, Gonzalo Torroba, and Sho Yaida for helpful discussions.  X.D., B.H., and E.S. are grateful to the KITP and the UCSB Department of Physics for an excellent year during which the major portion of this work was completed. A.W. would like to thank the KITP at UC Santa Barbara for their warm hospitality during the ``Strings at the LHC and in the Early Universe'' 2010 workshop, where part of this work was completed. The research of X.D., B.H., and E.S. was supported in part by the National Science Foundation under grant PHY05-51164, by NSF grant PHY-0244728, and by the DOE under contract DE-AC03-76SF00515.  B.H. is also supported by a William K. Bowes Jr. Stanford Graduate Fellowship.  The research of A.W. was supported in part by the Alexander von Humboldt Foundation, by NSF grant PHY-0244728, as well as by the ``Impuls und Vernetzungsfond'' of the Helmholtz Association of German Research Centres under grant HZ-NG-603.

\begingroup\raggedright\endgroup

\end{document}